\documentclass[
amsmath, amssymb, 
aps, pra, reprint,
twocolumn,
superscriptaddress,
floatfix,
showkeys
]{revtex4-1}

\usepackage{graphicx} 
\usepackage{float}
\usepackage{wrapfig}
\usepackage{array}
\usepackage{diagbox}
\usepackage{chngpage}
\usepackage{subfigure}
\usepackage{multirow}
\usepackage{tabularx}
\usepackage{hhline}
\usepackage[table]{xcolor}

\usepackage{braket}
\usepackage{accents}
\usepackage{upgreek}
\usepackage{color}

\usepackage{csquotes}
\usepackage{datetime}
\usepackage{units}
\usepackage{siunitx}

\usepackage{appendix}
\usepackage{mathtools}

\usepackage{textcomp}

\newcommand{\Yb}{\ensuremath{{}^{171}\textrm{Yb}^{+}}}
\newcommand{\ion}[2]{\ensuremath{{}^{#2}\textrm{#1}^{+}} }
\newcommand{\LS}[2]{\ensuremath{{}^2{\textrm{#1}}_{#2}}}
\newcommand{\LSB}[3]{\ensuremath{{}^{#1}{\textrm{#2}}[#3]_{#3}}}

\begin{document}

\title{Precision Characterization of the $^2$D$_{5/2}$ State and\\Quadratic Zeeman Coefficient in \Yb}

\author{T. R. Tan}
\thanks{tingrei86@gmail.com}
	\affiliation{
	The University of Sydney, School of Physics, NSW, 2006, Australia }
\author{C. L. Edmunds}
    \affiliation{
	ARC Centre of Excellence for Engineered Quantum Systems, The University of Sydney, School of Physics, NSW, 2006, Australia }
\author{A. R. Milne}
	\affiliation{
	ARC Centre of Excellence for Engineered Quantum Systems, The University of Sydney, School of Physics, NSW, 2006, Australia }
\author{M. J. Biercuk}
	\affiliation{
	ARC Centre of Excellence for Engineered Quantum Systems, The University of Sydney, School of Physics, NSW, 2006, Australia }
\author{C. Hempel}
\thanks{cornelius.hempel@gmail.com}
	\affiliation{
	ARC Centre of Excellence for Engineered Quantum Systems, The University of Sydney, School of Physics, NSW, 2006, Australia }
	\affiliation{
    The University of Sydney Nano Institute (Sydney Nano), The University of Sydney, NSW 2006, Australia	}

\date{\today}

\begin{abstract}
We report measurements of the branching fraction, hyperfine constant, and second-order Zeeman coefficient of the D$_{5/2}$ level in $^{171}$Yb$^+$ with up to two orders-of-magnitude improvement in precision compared to previously reported values. We estimate the electric quadrupole reduced matrix element of the S$_{1/2}$ $\leftrightarrow$ D$_{5/2}$ transition to be 12.5(4) $e a_0^2$. Furthermore, we determine the transition frequency of the F$_{7/2}$ $\leftrightarrow$ $^{1}$D$[3/2]_{3/2}$ at 760~nm with a $\sim$25-fold improvement in precision. These measurements provide benchmarks for quantum-many-body atomic-physics calculations and provide valuable data for efforts to improve  quantum information processors based on \Yb. 
\end{abstract}

\maketitle

Trapped atomic ions combine the benefits of long storage time and excellent isolation from the environment, providing an attractive physical system for numerous applications in quantum technology \cite{Wineland:1998,Knoop:2009}. Among the singly-charged ions most commonly investigated, \Yb~has found applications in tests of fundamental physics~\cite{Flambaum.2018, Counts:2020}, frequency metrology~\cite{Fisk.1997, Huntemann:2016}, and quantum information processing (QIP)~\mbox{\cite{Olmschenck:2007, Soare.2014, Mavadia:2018, Monroe.2019}}. All of these applications rely on a deep understanding of atomic structure theory, which is guided by and refined through precision measurements~\cite{Fawcett:1991,Itano:2006,Sahoo:2011,Dzuba:2011,Porsev:2012,Gossel:2013,Roberts:2014,Nandy:2014}. However, accurate calculation of the atomic properties for \Yb~has proven to be highly challenging due to its large number of valence electrons and complications arising from mixing between electronic configurations~\cite{Porsev:2012}, making it a particularly interesting system for experimental investigation. 

\begin{figure}
    \centering
    \includegraphics[width=1\linewidth]{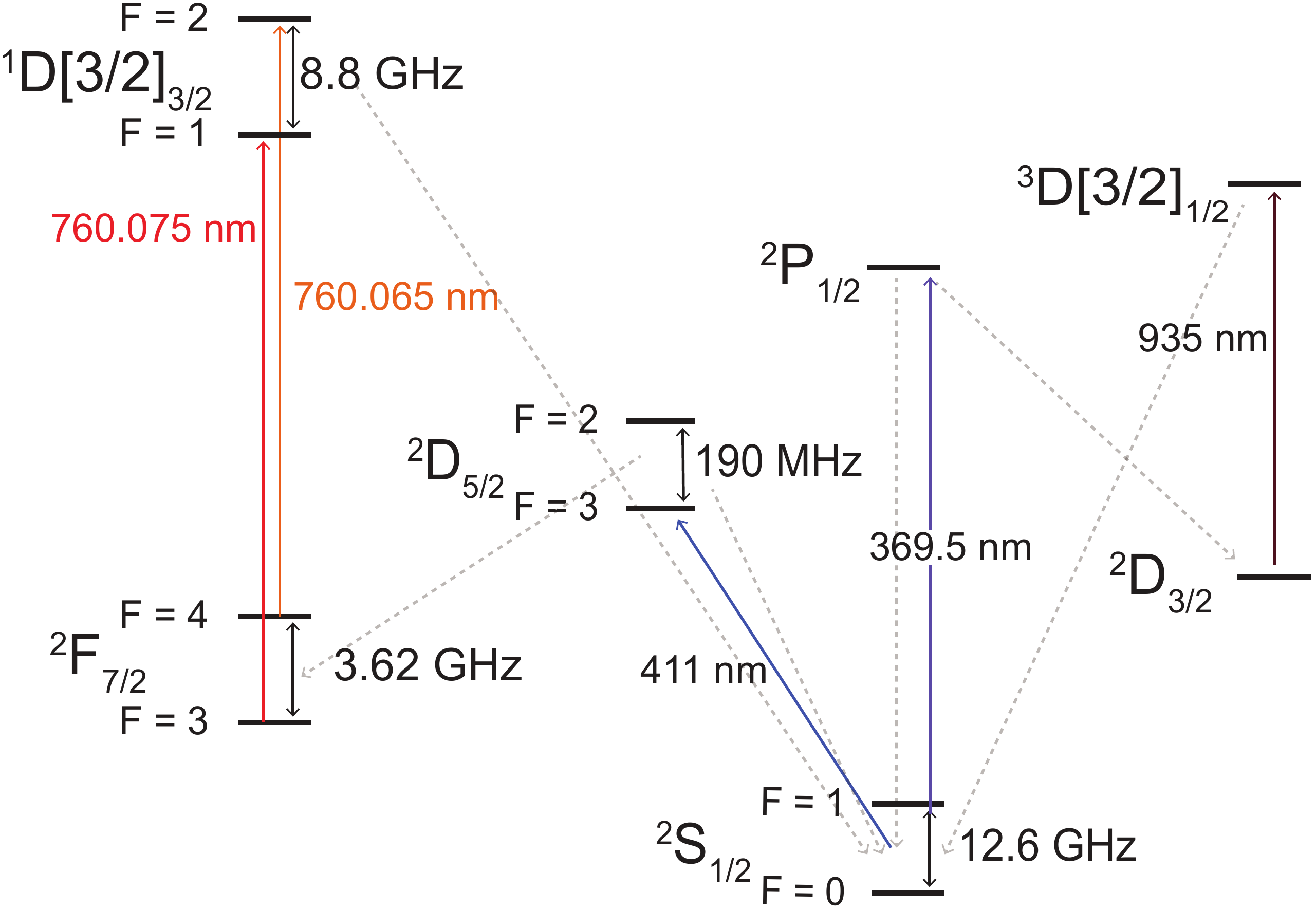}
    \caption{Selected energy levels and transition  wavelengths in \Yb. Dashed lines show relevant decay channels. Our work focuses on the 411~nm electric quadrupole (E2) $\LS{S}{1/2} \leftrightarrow \LS{D}{5/2}$ transition. The $\LS{D}{5/2}\ket{F=3}$ decays to $\LS{S}{1/2}\ket{F=1}$ and $\LS{F}{7/2}\ket{F=3,4}$ and $\LS{D}{5/2}\ket{F=2}$ decays to $\LS{S}{1/2}\ket{F=0,1}$ and $\LS{F}{7/2}\ket{F=3}$. A 760~nm laser is used for fast repumping of $\LS{F}{7/2}$ to $\LS{S}{1/2}$ via $\LSB{1}{D}{3/2}$. Some hyperfine structures and Zeeman sublevels are omitted for clarity. }
    \label{fig:full_energy_levels}
\end{figure}

The nuclear spin of \Yb is one half, causing the $\LS{S}{1/2}$ electronic ground state to split into a hyperfine doublet separated by $\sim$12.64~GHz. Its metastable levels (Fig.~\ref{fig:full_energy_levels}) include $\LS{D}{3/2}$, $\LS{D}{5/2}$, and $\LS{F}{7/2}$ with lifetimes of 52.7~ms~\cite{Yu:2000}, 7.2~ms~\cite{Taylor:1997}, and $\gtrsim$ 5.4~years~\cite{Roberts:2000}, respectively. The short lifetime of $\LS{D}{5/2}$ has made the state less attractive for frequency metrology compared to $\LS{D}{3/2}$~\cite{Webster.2010, Tamm:2014, Schacht.2015} and $\LS{F}{7/2}$~\cite{Hosaka:2009,Huntemann:2012,Huntemann:2016}; experimental measurements on $\LS{D}{5/2}$ were last reported more than 20 years ago~\cite{Roberts:1999}. Still, measurements with improved precision provide a good benchmark to verify challenging and model-dependent atomic physics calculations. Furthermore, the metastable ${}^2$D levels in \Yb provide extra degrees of freedom for applications in QIP. These find use in implementing entangling gate operations~\cite{Baldwin:2020} or to scalably improve qubit measurement fidelities with minimal technical overhead~\cite{Edmunds.2020b}.

In this Letter, we report precision measurements on branching ratios, the electric quadrupole (E2) reduced matrix element, the quadratic Zeeman (QZ) coefficient, and the hyperfine splitting of the $\LS{D}{5/2}$ state in \Yb, with precision up to two order of magnitude higher than the previous best reported values. Our spectroscopic experiments use a single ion in a Paul trap with a storage lifetime of several months, and a new stabilized solid-state laser near 411~nm.  Furthermore, we provide a more precise measurement of the transition between \mbox{$\LS{F}{7/2}$ $\leftrightarrow$ $\LSB{1}{D}{3/2}$} at 760~nm than previously reported~\cite{Sugiyama:1999,Jaua:2015, Mulholland:2019}. 

\begin{figure}
    \centering
    \includegraphics[width=1\linewidth]{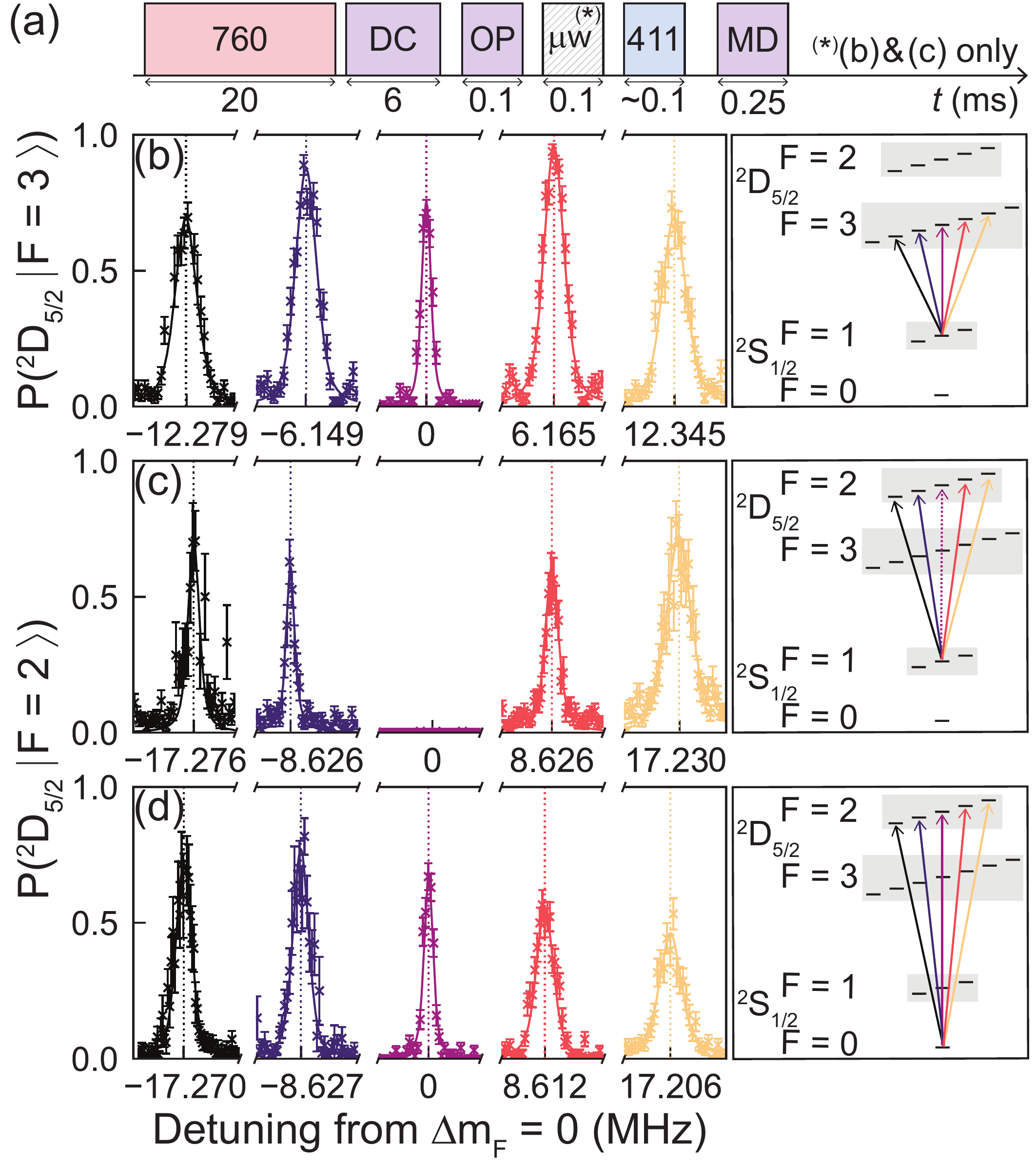}
    \caption{Spectra of the $\LS{S}{1/2} \leftrightarrow \LS{D}{5/2}$ transitions at 411~nm. (a) The spectroscopy pulse sequence including 760~nm repumping, Dopper cooling (DC), optical pumping to prepare $\LS{S}{1/2}\ket{F=0}$ (OP), microwave excitation to $\LS{S}{1/2}\ket{F=1,m_F=0}$ in select cases ($\upmu$w), the 411~nm spectroscopy pulse and detection (MD), which includes the same frequency components as DC but is tuned closer to resonance. (b,c,d) 411 nm laser frequency scans around the line center at $\Delta m_F=0$ for the different hyperfine transitions indicated in corresponding colors on the right. In (c), the $\Delta m_F=0$ transition cannot be driven due to atomic selection rules. Error bars are calculated from quantum projection noise (QPN) and the data are fitted to Voigt profiles. The full width at half maximum of the lines varies between 3 and 11~kHz, with the largest impact arising from laser polarization-dependent coupling strength differences across different transitions. The precision of the indicated center frequencies is limited by the fitting error to (0.1-2)~kHz; amplitude differences are due to polarization and beam orientation with respect to the quantization axis. }
    \label{fig:411_spectra}
\end{figure}

The experimental setup is detailed in Ref.~\cite{Edmunds:2020}. A permanent magnet produces a quantization field of $\sim$0.44~mT at the ion position, which we infer using the 13.98(1)~GHz/T linear Zeeman shift of the $\LS{S}{1/2}$ stretched states~\cite{Meggers:1967,Ludlow:2015}. A 369.5~nm laser in conjunction with electro-optic modulators (EOMs) is used for Doppler cooling, state preparation, and detection. Separate 369.5~nm laser beamlines allow for selective measurements of just the population in $\LS{S}{1/2}\ket{F=1}$ using the \enquote{standard detection} (SD), or the entire $\LS{S}{1/2}$ \enquote{manifold detection} (MD). A 935~nm laser continuously repumps population which has decayed into the $\LS{D}{3/2}$ manifold. 

We drive the \mbox{$\LS{S}{1/2}$ $\leftrightarrow$ $\LS{D}{5/2}$} transition using an external-cavity diode laser at 411~nm, which is locked to an ultra-low expansion reference cavity with a free spectral range (FSR) of 1.5~GHz, a finesse of approximately 32,000, and a frequency drift rate of $\sim$ 320~mHz/s. An additional, cavity-stabilized 760~nm diode laser configured with a cateye reflector~\cite{MOGCatEye,Thompson:2012} is used to provide efficient repumping from the long-lived $\LS{F}{7/2}$ via ${}^3$D[3/2]$_{1/2}$, which has a lifetime of $\sim$~28.6~ns and decays primarily to the $\LS{S}{1/2}$ ground states~\cite{Berends:1993,Sugiyama:1999}. To maximise repumping efficiency, an EOM driven at 5.258~GHz is used to create an additional laser frequency component, matching the hyperfine splittings of 3.620~GHz~\cite{Webster:1999,Taylor:1999} and 8.8 GHz~\cite{Mulholland:2019} for $\LS{F}{7/2}$ and $\LSB{1}{D}{3/2}$, respectively. The 5.258~GHz microwave signal used to drive the EOM is tuned by maximizing fluorescence counts on the 369.5~nm detection transition while simultaneously illuminating the ion with 411~nm and 760~nm light. With an intensity of $\sim$ 0.63~\nicefrac{W}{mm$^2$}, the typical repumping time from the $\LS{F}{7/2}$ state to the $\LS{S}{1/2}$ manifold is $\sim$20 ms, significantly faster than other repumping channels~\cite{Gill:1995,Taylor:1997,Sugiyama:1999, Yu:2000}, such as the 638~nm transition to $^1$D[5/2]$_{5/2}$. All laser beams are controlled by acousto-optic modulators driven by direct digital synthesis sources referenced to a rubidium frequency standard. 

\begin{figure}
    \centering
    \includegraphics[width=1\linewidth]{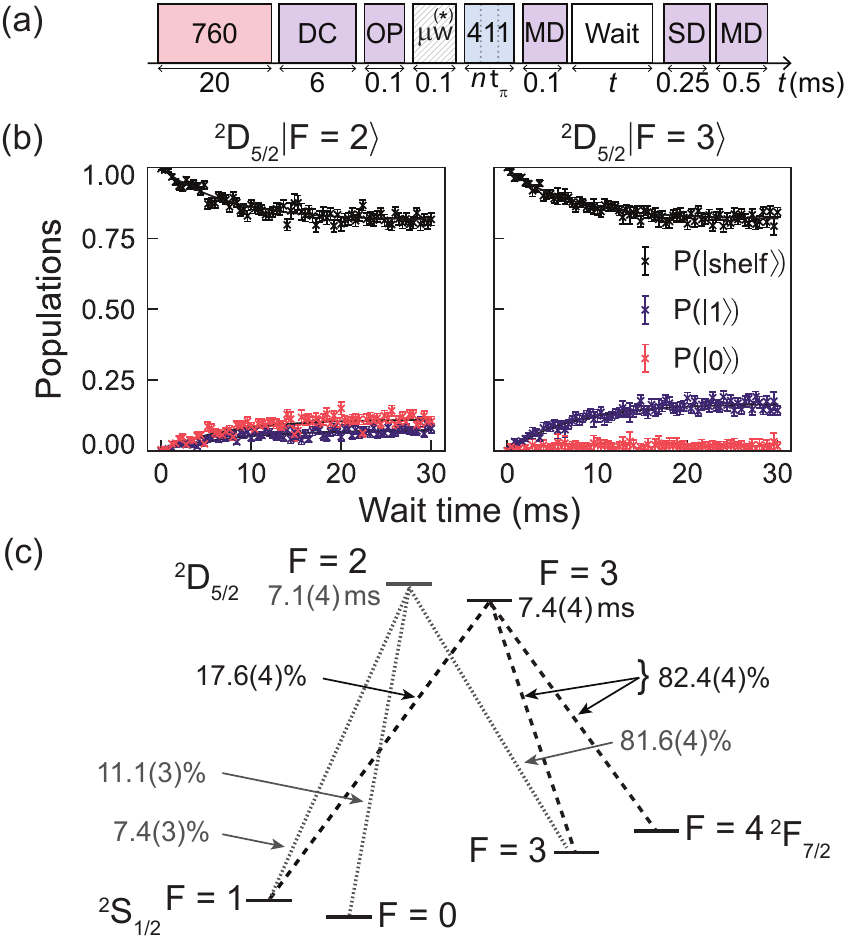}
    \caption{Lifetime and branching ratio measurements. (a) Measurement pulse sequence similar to Fig.~\ref{fig:411_spectra}(a), with an added variable wait time $t$, surrounded by MD and SD pulses for post selection and measurement of branching to the two $\LS{S}{1/2}$ levels, respectively. A series of $n=5$ $\pi$ pulses on the accessible 411~nm transitions to different Zeeman sublevels maximizes the shelving success to $\LS{D}{5/2}\ket{F=2}$. Due to the similar Zeeman splitting in $\LS{S}{1/2}$ $\ket{F = 1}$ and $\LS{D}{5/2}$ $\ket{F = 3}$ manifolds, a similar $n=3$ procedure is used to prepare $\LS{D}{5/2}\ket{F=3}$ from $\LS{S}{1/2}\ket{F=1,m_F=0}$. (b) State decay curves as a function of wait time after initialization of the target state. The populations P$(\ket{0})$ and P$(\ket{1})$ refer to the two hyperfine levels in the $\LS{S}{1/2}$ ground state manifold, while P$(\ket{\textrm{shelf}})$ refers to the shelved population in $^2$D$_{5/2}$ or $^2$F$_{7/2}$. (c) Overview of the decay channels and branching fractions measured in this work. }
    \label{fig:D5half_decays}
\end{figure}

We begin by performing spectroscopy on the \mbox{$\LS{S}{1/2} \leftrightarrow \LS{D}{5/2}$} optical transition near 411~nm, resolving individual hyperfine and Zeeman levels as shown in Figure~\ref{fig:411_spectra}. The ion is initialized in $\LS{S}{1/2}\ket{F=0}$ by optical pumping and optionally transferred to $\LS{S}{1/2}\ket{F=1, m_F=0}$ through application of a microwave $\pi$-pulse. After applying a 411~nm spectroscopy pulse of fixed duration, state-dependent fluorescence is induced by a 369.5~nm (MD) pulse that distinguishes population in the $\LS{S}{1/2}$ manifold (``bright'') from that in other levels (``dark''). A 20~ms pulse at 760~nm is applied to repump population decayed to $\LS{F}{7/2}$ at the beginning of each measurement cycle. The center frequency for each transition is extracted by fitting the spectroscopic peaks to Voigt profiles. We measure a frequency of 729.487\thinspace752(178)~THz for the $\LS{S}{1/2}\ket{F=0} \leftrightarrow \LS{D}{5/2}\ket{F=2, m_F=0}$ transition where the uncertainty corresponds to $\pm 0.1$ pm. We take this conservative value from our wavemeter's maximum guaranteed accuracy when the measurement wavelength is more than $\pm$ 200~nm away from the calibration wavelength at 632~nm. Our frequency measurement is consistent with previous measurements~\cite{Roberts:1999}. We note that a more precise measurement of the $\LS{S}{1/2}$ $\leftrightarrow$ $\LS{D}{5/2}$ transition in the $^{172}$Yb$^+$ isotope has recently been reported~\cite{Fuerst:2020}.

\begin{table}
\begin{tabularx}{\columnwidth}{ 
   >{\centering\arraybackslash}l
   >{\raggedright\arraybackslash}X
   >{\raggedright\arraybackslash}X 
   >{\raggedright\arraybackslash}X }
     \hhline{====}
     \parbox[c][30pt][c]{\textwidth/7}{}
     & \parbox[c][30pt][c]{\textwidth/13}
     {Decay\\to $\LS{S}{1/2}$}
     & \parbox[c][30pt][c]{\textwidth/13}
     {Decay\\to $\LS{F}{7/2}$}
     & \parbox[c][30pt][c]{\textwidth/15}
     {Lifetime (ms)} 
     \\ \hline
     \parbox[c][30pt][c]{\textwidth/7}
     {This work (exp.) $\Yb\ket{F=3}$}
     & 17.6(4)\%
     & 82.4(4)\%
     & \,\, 7.1(4)
     \\
     \parbox[c][30pt][c]{\textwidth/7}
     {This work (exp.) $\Yb\ket{F=2}$ }
     & 18.4(4)\%
     & 81.6(4)\%
     & \,\, 7.4(4)
     \\
     \hline
     \parbox[c][30pt][c]{\textwidth/6}
     {Taylor et al.~\cite{Taylor:1997} (1997 exp.) $\ion{Yb}{172}$}
     & 17(4)\%
     & 83(3)\%
     & \,\, 7.2(3)
     \\
     
     \parbox[c][30pt][c]{\textwidth/6}
     {Yu et al.~\cite{Yu:2000}\\ (2000 exp.) $\ion{Yb}{174}$}
     & \multicolumn{1}{c}{-}
     & \multicolumn{1}{c}{-}
     & \,\, 7.0(4)
     \\ \hline
     
     \parbox[c][30pt][c]{\textwidth/7}
     {Fawcett et al.~\cite{Fawcett:1991} (1991 calc.) Yb II}
     & 19.7\%
     & 80.3\%
     & \,\, 5.74
     \\ \hhline{====}
\end{tabularx}
\caption{Branching ratios and lifetimes for the $\LS{D}{5/2}\ket{F=2,3}$ state contrasted with previous measurements and a theoretical prediction. Uncertainties of $1\sigma$ are derived from exponential fits to the state evolution shown in Fig.~\ref{fig:D5half_decays}.}
\label{tab:D5half_decays}
\end{table}

The lifetimes and decay branching ratios of $\LS{D}{5/2}$ are measured by a sequence that initializes in $\LS{D}{5/2}\ket{F=2}$ or $\LS{D}{5/2}\ket{F=3}$ through multiple sequential $\pi$ pulses to different Zeeman sublevels, followed by an immediate detection used for post selection on events with successful transfer, see Fig.~\ref{fig:D5half_decays}(a). Following a variable wait time, two final detection pulses determine the population in the $\LS{S}{1/2}$ states and shelved manifolds. The lifetime is extracted by fitting exponential decays, weighted by QPN, to time-delayed data (Fig.~\ref{fig:D5half_decays}(b,c)), where we take the asymptotic limits of the fit as branching ratios. This is the first $\LS{D}{5/2}$ lifetime and branching fraction measurement made on \Yb. The values agree with previous measurements on different isotopes which we summarize in Table~\ref{tab:D5half_decays}. The asymptotic limits for the $\ket{F=0}$ and $\ket{F=1}$ populations are 11.1(3)\% and 7.4(3)\%, which agree well with the expected relative decay branching fraction from $\LS{D}{5/2}\ket{F=2, m_F=0}$ to $\LS{S}{1/2}\ket{F=1}$ and $\LS{D}{5/2}\ket{F=2, m_F=0}$ to $\LS{S}{1/2}\ket{F=0}$. From the lifetime and branching ratios, we estimate the E2 reduced matrix element to be 12.5(4) $e a_0^2$, with $e$ and $a_0$ the elementary charge and Bohr radius, respectively, which is consistent with theoretical predictions~\cite{Martin:2002,Sahoo:2011}.

\begin{figure}
    \centering
    \includegraphics[width=1\linewidth]{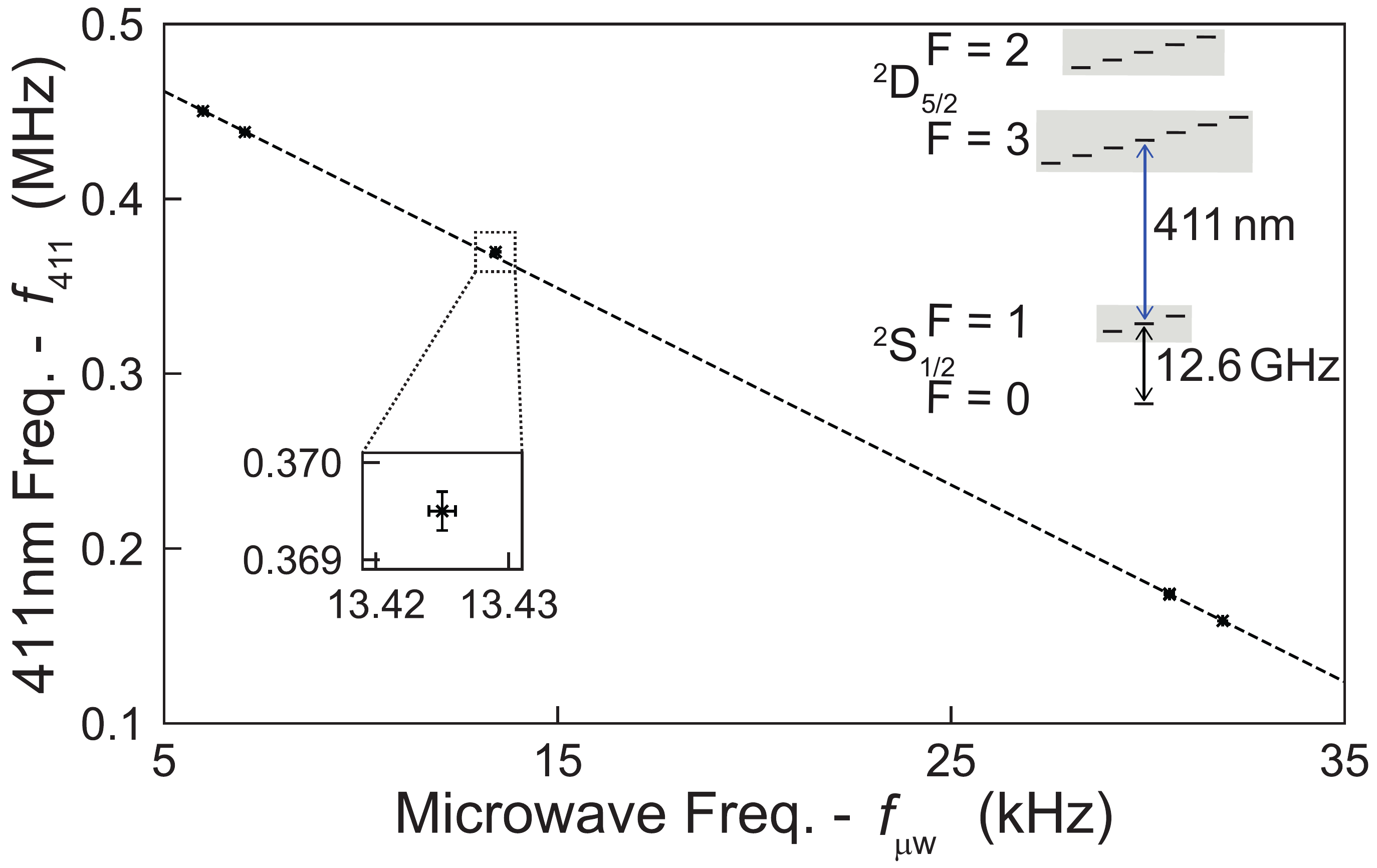}
    \caption{Frequency of $\LS{D}{5/2}\ket{F=3, m_F=0}$ near 411~nm vs. frequency of $\LS{S}{1/2}\ket{F=1, m_F=0}$ with respect to $\LS{S}{1/2}\ket{F=0}$ at different magnetic field strengths. They are deduced from the $\LS{S}{1/2}\ket{F=1,m_F=1}$ $\leftrightarrow$ $\LS{D}{5/2}\ket{F=3, m_F=0}$ and the $\LS{S}{1/2}\ket{F=0}$ $\leftrightarrow$ $\LS{S}{1/2}\ket{F=1, m_F=0}$ transitions which are illustrated on the right.
    Typical statistical uncertainties of the frequency measurements are approximately 200~Hz and 0.2~Hz for the optical and microwave frequencies, respectively, with the error bar size highlighted in the lower inset. The frequency offsets are $f_{\mu w} = 12.642\thinspace812\thinspace118\thinspace466$~GHz, and $f_{411} = 729.487\thinspace559$~THz. We deduce a slope  of -11.27(4) where the uncertainty is the standard error of the fit.}
    \label{fig:SecondOrderZeemanData}
\end{figure}

To deduce the second-order Zeeman coefficient of the $\LS{D}{5/2}\ket{F=3, m_F=0}$ state, we vary the magnetic field at the ion location over an interval of $\sim$ [0.44,1] mT through adjustment of a permanent magnet's placement. For each B-field configuration, the frequencies of the $\LS{S}{1/2}\ket{F=0} \leftrightarrow \LS{S}{1/2}\ket{F=1, m_F=0}$ hyperfine transition at $\sim$12.6~GHz and the $\LS{S}{1/2}\ket{F=1,m_F=0} \leftrightarrow \LS{D}{5/2}\ket{F=3, m_F=0}$ transition at 411~nm are recorded, updated, and averaged through interleaved Ramsey interrogations using wait times of 50~ms and 0.1~ms, respectively. We combine the microwave and optical frequency measurements to find the $\LS{D}{5/2}\ket{F=3, m_F=0}$ frequency with respect to the $\LS{S}{1/2}\ket{F=0}$ ground state. In Fig.~\ref{fig:SecondOrderZeemanData}, we plot the relative frequency shifts of $\LS{S}{1/2}\ket{F=0}$ $\leftrightarrow$ $\LS{D}{5/2}\ket{F=3, m_F=0}$ and $\LS{S}{1/2}\ket{F=0}$ $\leftrightarrow$ $\LS{S}{1/2}\ket{F=1, m_F=0}$ after compensating the measured optical frequencies for cavity drift. We determine the ratio of the quadratic Zeeman coefficients to be -11.27(4). 
Taken together with the often quoted value of \mbox{0.03108~Hz/$\upmu$T$^2$}~\cite{Vanier:1989} for the quadratic Zeeman coefficient in the $\LS{S}{1/2}$ ground state transition, this yields a QZ coefficient of \mbox{$-0.350(1$)~Hz/$\upmu$T$^{2}$} for $\LS{D}{5/2}\ket{F=3,m_F=0}$ with respect to the $\LS{S}{1/2}\ket{F=0}$ ground state. The quadratic Zeeman coefficient of $\LS{D}{5/2}\ket{F=3,m_F=0}$ is 
\begin{equation}
\frac{1}{4}\frac{(g_{J,D}-g_{I})^2 \mu_B^2 }{h^2 \Delta_{HF,D}},
\end{equation}
which has the same magnitude but opposite sign as $\LS{D}{5/2}\ket{F=2,m_F=0}$ \cite{Itano:2000}; here, $\Delta_{HF,D}$, $g_{J,D}$, and $\mu_B$ are the hyperfine splitting, Land\'{e} $g$-factor of $\LS{D}{5/2}$, and Bohr's magneton, respectively. The nuclear $g$-factor of Yb$^+$ is $g_I=-5.38 \times 10^{-4}$ \cite{Stone:2015}. The absolute value of our measurement is an order of magnitude more precise than the best published result of \mbox{0.38(8)~Hz/$\upmu$T$^{2}$} measured on $\LS{S}{1/2}\ket{F=0} \leftrightarrow \LS{D}{5/2}\ket{F=2, m_F=0}$~\cite{Roberts:1999}. We further quantified a potential AC Zeeman shift in the transition frequencies due to the trap's radio-frequency field using an Autler-Townes spectroscopy method first demonstrated in Ref. \cite{Gan:2018}. The magnitude of the B-field perpendicular to the quantization axis oscillating at the trapping radio frequency is measured to be less than 10~nT, having a negligible impact on our measurements. 

The quadratic Zeeman coefficient of the ground state transition is
\begin{equation}
QZ_S=\frac{1}{2}\frac{(g_{J,S}-g_I)^2 \mu_B^2}{h^2 \Delta_{HF,S}}
\label{Eq:HFS}
\end{equation}
where $\Delta_{HF,S} = 12.642\thinspace812\thinspace118\thinspace466$~GHz \cite{Fisk.1997} is the hyperfine splitting of in the $\LS{S}{1/2}$ ground state and $g_{J,S}$ is Land\'{e} $g$-factor for $\LS{S}{1/2}$. The quadratic Zeeman coefficient of $\LS{D}{5/2}\ket{F=3,m_F=0}$ with respect to the $\LS{S}{1/2}\ket{F=0}$ ground state is 

\begin{equation}
QZ_{D,F=3}=\frac{1}{4}\frac{(g_{J,D}-g_{I})^2 \mu_B^2 }{h^2 \Delta_{HF,D}} - 
\left(-\frac{1}{4}\frac{(g_{J,S}-g_I)^2 \mu_B^2 }{h^2 \Delta_{HF,S}}\right).
\label{Eq:HFD}
\end{equation}

A recent theory calculation of $g$-factor anomalies in \Yb gives $g_{J,S}$ = 2.0031(3) and $g_{J,D}$ = 1.20051(2) where the uncertainties are the estimated maximum error of the calculation \cite{Gossel:2013}. Using these values, we evaluate the QZ coefficients to $QZ_S$ =  \mbox{0.03110(1)~Hz/$\upmu$T$^2$} and  $QZ_{D,F=3}$ = \mbox{$-0.35606(1)$~Hz/$\upmu$T$^2$}, yielding a ratio of the quadratic Zeeman coefficient of -11.448(2), which does not agree with our measurement in Fig.~\ref{fig:SecondOrderZeemanData}. This could indicate an effect not captured by Eqs. \eqref{Eq:HFS} and \eqref{Eq:HFD}, or a discrepancy in the $g_J$ values above. Spectroscopic measurement values of $g_{J,S}$ and $g_{J,D}$ were given by Meggers as 1.998 and 1.202 \cite{Meggers:1967} without specifying uncertainties. Assuming an uncertainty of 0.002 for both values, we evaluate the QZ coefficients and ratio to be $QZ_S$ =  \mbox{0.03094(4)~Hz/$\upmu$T$^{2}$}, $QZ_{D,F=3}$ = -0.3571(12)~\mbox{Hz/$\upmu$T$^2$} and  -11.54(4), which also does not agree with our observations. Our measurements notwithstanding, we note that Meggers' ground state $g$-factor value of 1.998 implies a $g$-factor anomaly correction of $-4\times 10^{-3}$ which is of opposite sign to the $\delta g = +79.8\times 10^{-5}$ reported in Ref.~\cite{Gossel:2013}. This suggests potential issues in previous $g$-factor measurements in Yb$^+$. 

The hyperfine splitting of \LS{D}{5/2} can be deduced from the difference of the transition frequencies between $\LS{S}{1/2}\ket{F=1,m_F=0}$ and $\LS{D}{5/2}\ket{F=2,m_F=0}$ as well as $\LS{D}{5/2}\ket{F=3,m_F=0}$. Electric quadrupole transition selection rules forbid the $\LS{S}{1/2}\ket{F=1,m_F=0}$ $\leftrightarrow$ $\LS{D}{5/2}\ket{F=2,m_F=0}$ transition to be directly driven; therefore we infer its value from measurements of $\LS{S}{1/2}\ket{F=1,m_F=0}$ $\leftrightarrow$ $\LS{D}{5/2}\ket{F=2,m_F=\pm 1}$. The frequency of the 411~nm laser is stabilized to the same cavity mode throughout this measurement in order to avoid uncertainties from wavemeter's accuracy or cavity's FSR in the determination of hyperfine splitting. All transition frequencies are corrected for cavity drift over the measurement duration of $\sim 4$~ hours and QZ shifts at the B-field of 441.26(2)~$\upmu$T \cite{Itano:2000}. From our measurements we extract a hyperfine splitting of $\Delta_{HF,D} = -190.104(3)$~MHz and,  correspondingly, a hyperfine constant of \mbox{$A_{D, 5/2} = -63.368(1)$}~MHz. This measurement is more than two orders of magnitude more precise than the best previously reported value of $-191(2)$ MHz~\cite{Roberts:1999}, and limited by the precision of the QZ coefficients and fitting errors. Table~\ref{tab:HF_splitting} summarizes both measured and theoretical values for $A_{D, 5/2}$, illustrating the difficulty of performing exact calculations for \Yb. 

\renewcommand{\arraystretch}{1.5}
\begin{table}[b]
\begin{tabularx}{\columnwidth}{ 
l X l}
    \hline\hline
    Reference
& & {$A_{D, 5/2}$ (MHz)}
    \\ \hline
    This work & (2020 experiment) &  -63.368(1) \\ 
    Roberts et al.~\cite{Roberts:1999} &(1999 experiment) &  -63.6(7) \\
    \hline
    Nandy et al.~\cite{Nandy:2014} &(2014 calculation) & -69(6) \\
    Porsev et al.~\cite{Porsev:2012} &(2012 calculation) & -96 \\ 
    Sahoo et al.~\cite{Sahoo:2011} &(2011 calculation) & -48(15) \\
    Itano~\cite{Itano:2006} &(2006 calculation)  & -12.58 \\ 
    \hline\hline
\end{tabularx}
\caption{Comparison of measurements and calculations of the $\LS{D}{5/2}$ hyperfine constant.}
\label{tab:HF_splitting}
\end{table}

As a final measurement, we determine the $\LS{F}{7/2}$ $\leftrightarrow$ $\LSB{1}{D}{3/2}$ transition near 760~nm, used to repump population to $\LS{S}{1/2}$. We first prepare population in $\LS{D}{5/2}\ket{F=2,3,m_F=0}$ using a 411~nm laser pulse, followed by a 10~ms wait time to allow population decay to $\LS{F}{7/2}$. A subsequent brief application of the 369.5~nm MD beam allows us to remove instances of $\LS{S}{1/2}$ decay from the data in post processing. A 760~nm laser pulse is then applied to induce a transition to the short-lived $\LSB{1}{D}{3/2}$ level, which primarily decays to $\LS{S}{1/2}$. Successful repump events then yield fluorescence under application of the 369.5~nm MD light. The measured center frequencies are 
\begin{eqnarray}
\LS{F}{7/2}\ket{F\!=\!3} \leftrightarrow \LSB{1}{D}{3/2}\ket{F\!=\!1}&:& 394.424\thinspace943(20) \mathrm{THz},\nonumber\\
\LS{F}{7/2}\ket{F\!=\!4} \leftrightarrow \LSB{1}{D}{3/2}\ket{F\!=\!2}&:& 394.430\thinspace203(16) \mathrm{THz},\nonumber
\end{eqnarray}
respectively. Here the uncertainties are dominated by line broadening due to our inability to prepare the ion in a specific $\LS{F}{7/2}$ Zeeman level as well as the wavemeter's specified absolute accuracy, which is 10 MHz within $\pm$ 200~nm of the calibration wavelength of 632~nm. These measurements improve precision by 25-fold over previously reported values~\cite{Mulholland:2019}.

In conclusion, we report measurements of the hyperfine splitting, branching ratios, and quadractic Zeeman coefficient of $\LS{D}{5/2}$ in \Yb with improved precision and investigate the 760~nm transition to efficiently depopulate the  $\LS{F}{7/2}$ state. These new results can be used to benchmark and verify quantum many-body calculations in Yb$^+$, e.g., in support of studies of PNC physics and the related nuclear anapole moment \cite{Sahoo:2011,Dzuba:2011,Porsev:2012,Gossel:2013,Roberts:2014}. Furthermore, these results provide the necessary characterizations required for improving the measurement fidelity of \Yb hyperfine qubits as detailed in our separate manuscript Ref.~\cite{Edmunds.2020b}.

\section*{Acknowledgements}
T.R. Tan acknowledges support from the Sydney Quantum Academy. The authors thank A. Singh and T.F. Wohlers-Reichel for their contribution to assembling and maintenance of the experimental setup. This work was partially supported by the Intelligence Advanced Research Projects Activity Grant No. W911NF-16-1-0070, the US Army Research Office Grant No. W911NF-14-1-0682, the Australian Research Council Centre of Excellence for Engineered Quantum Systems Grant No. CE170100009, and a private grant from H.\&A. Harley. \newline

\noindent TRT and CE contributed equally to this work.


\begin{thebibliography}{99}

\bibitem{Wineland:1998}
D. J. Wineland, C. Monroe, W. M. Itano, D. Leibfried, B. E. King, and D. M. Meekhof, ``Experimental issues in coherent quantum-state manipulation of trapped atomic ions,'' Journal of Research of the National Institute of Standards and Technology, {\bf 103}, 259 (1998).

\bibitem{Knoop:2009}
M. Knoop, L. Hilico, and J. Eschner, ``Modern applications of trapped ions,'' Journal of Physics B: Atomic, Molecular and Optical Physics, {\bf 42}, 150201 (2009).

\bibitem{Flambaum.2018}
V. V. Flambaum, A. J. Geddes, and A. V. Viatkina, ``Isotope shift, nonlinearity of King plots, and the search for new particles,'' Phys. Rev. A {\bf 97}, 032510 (2018).

\bibitem{Counts:2020}
I. Counts, J. Hur, D. P. L. Aude Craik, H. Jeon, C. Leung, J. C. Berengut, A. Geddes,  A. Kawasaki, W. Jhe, V. Vuleti\'c, ``Evidence for Nonlinear Isotope Shift in ${\mathrm{Yb}}^{+}$ Search for New Boson,'' Phys. Rev. Lett. {\bf 125}, 123002 (2020).

\bibitem{Huntemann:2016}
N. Huntemann, C. Sanner, B. Lipphardt, Chr. Tamm, and E. Peik, ``Single-Ion Atomic Clock with $3\times 10^{-18}$ Systematic Uncertainty,'' Phys. Rev. Lett. {\bf 116}, 063001 (2016).

\bibitem{Fisk.1997}
P. T. H. Fisk, M. J. Sellars, M. A Lawn, and G. Coles, ``Accurate measurement of the 12.6 GHz "clock" transition in trapped $^{171}$Yb$^+$ ions,'' IEEE Transactions on Ultrasonics, Ferroelectrics and Frequency Control {\bf 44}, 344 (1997).

\bibitem{Soare.2014}
A. Soare, H. Ball, H. D. Hayes, D. J. Sastrawan, M. C. Jarratt, J. J. McLoughlin, X. Zhen, T. J. Green, and M. J. Biercuk, ``Experimental noise filtering by quantum control,'' Nature Physics {\bf 10}, 825 (2014).

\bibitem{Olmschenck:2007}
S. Olmschenk, K. C. Younge, D. L. Moehring, D. Matsukevich, P. Maunz, and C. Monroe, ``Manipulation and detection of a trapped ${\mathrm{Yb}}^{+}$ hyperfine qubit,'' Phys. Rev. A {\bf 76}, 052314 (2007).

\bibitem{Mavadia:2018}
S. Mavadia, C. L. Edmunds, C. Hempel, H. Ball, F. Roy, T. M. Stace, and M. J. Biercuk, ``Experimental quantum verification in the presence of temporally correlated noise,'' NPJ Quantum Inf. {\bf 4}, 7 (2018).

\bibitem{Monroe.2019}
C. Monroe, W. C. Campbell, L.-M. Duan, Z.-X. Gong, A. V. Gorshkov, P. Hess, R. Islam, K. Kim, G. Pagano, P. Richerme, C. Senko, C and N. Y. Yao, ``Programmable Quantum Simulations of Spin Systems with Trapped Ions,'' Rev. Mod. Phys. {\bf 93}, 025001 (2021).

\bibitem{Fawcett:1991}
B. C. Fawcett and M. Wilson, ``Computed oscillator strengths, Land{\'e} g values, and lifetimes in Yb II,'' Atomic Data and Nuclear Data Tables {\bf 47}, 241 (1991).

\bibitem{Itano:2006}
Wayne M. Itano, ``Quadrupole moments and hyperfine constants of metastable states of ${\mathrm{Ca}}^{+}$, ${\mathrm{Sr}}^{+}$, ${\mathrm{Ba}}^{+}$, ${\mathrm{Yb}}^{+}$, ${\mathrm{Hg}}^{+}$, and Au,'' Phys. Rev. A {\bf 73}, 022510 (2006).

\bibitem{Sahoo:2011}
B. K. Sahoo and B. P. Das, ``Parity nonconservation in ytterbium ion,'' Phys. Rev. A {\bf 84}, 010502(R) (2011).

\bibitem{Dzuba:2011}
V. A. Dzuba and V. V. Flambaum ``Calculation of nuclear-spin-dependent parity nonconservation in $s$--$d$ transitions of Ba${}^{+}$,'' Phys. Rev. A {\bf 83}, 052513(R) (2011).

\bibitem{Porsev:2012}
S. G. Porsev, M. S. Safronova, and M. G. Kozlov, ``Correlation effects in $\textrm{Yb}^{+}$ and implications for parity violation,'' Phys. Rev. A {\bf 86}, 022504 (2012).

\bibitem{Gossel:2013}
G. H. Gossel, V. A. Dzuba, and V. V. Flambaum, ``Calculation of strongly forbidden M1 transitions and g-factor anomalies in atoms considered for parity-nonconservation measurements,'' Phys. Rev. A {\bf 88}, 034501 (2013).

\bibitem{Nandy:2014}
D. K. Nandy and B. K. Sahoo, ``Quadrupole shifts for the $^{171}\mathrm{Yb}^{+}$ ion clocks: Experiments versus theories,'' Phys. Rev. A {\bf 90}, 050503(R) (2014).

\bibitem{Roberts:2014}
B. M. Roberts, V. A. Dzuba, and V. V. Flambaum, ``Nuclear-spin-dependent parity nonconservation in s-d5/2 and s-d3/2 transitions,'' Phys. Rev. A {\bf 89}, 012502 (2014).

\bibitem{Yu:2000}
N. Yu, and L. Maleki, ``Lifetime measurements of the ${4f}^{14}5d$ metastable states in single ytterbium ions,'' Phys. Rev. A {\bf 61}, 022507 (2000).

\bibitem{Taylor:1997}
P. Taylor, M. Roberts, S. V. Gateva-Kostova, R. B. M. Clarke, G. P. Barwood, W. R. C. Rowley, and P. Gill, ``Investigation of the ${}^{2}{S}_{1/2}{\ensuremath{-}}^{2}{D}_{5/2}$ clock transition in a single ytterbium ion,'' Phys. Rev. A {\bf 56}, 2699 (1997).

\bibitem{Roberts:2000}
M. Roberts, P. Taylor, G. P. Barwood, W. R. C. Rowley, and P. Gill, ``Observation of the 2S1/2-2F7/2 electric octupole transition in a single 171Yb+ ion,'' Phys. Rev. A {\bf 62}, 020501(R) (2000).

\bibitem{Webster.2010}
S. Webster, R. Godun, S. King, G. Huang, B. Walton, V. Tsatourian, H. Margolis, S. Lea, and P. Gill, ``Frequency measurement of the $^2S_{1/2}$ $^2D_{3/2}$ electric quadrupole transition in a single \Yb ion,'' IEEE Transactions on Ultrasonics, Ferroelectrics and Frequency Control {\bf 57}, 592 (2010).

\bibitem{Tamm:2014}
Chr. Tamm, N. Huntemann, B. Lipphardt, V. Gerginov, N. Nemitz, M. Kazda, S. Weyers, and E. Peik, ``Cs-based optical frequency measurement using cross-linked optical and microwave oscillators,'' Phys. Rev. A {\bf 89}, 023820 (2014).

\bibitem{Schacht.2015}
M. Schacht, J. R. Danielson, S. Rahaman, J. R. Torgerson, J. Zhang, and M. M. Schauer, M M, ``171Yb+ 5D3/2 hyperfine state detection and F = 2 lifetime,'' Journal of Physics B: Atomic, Molecular and Optical Physics, {\bf 48} 065003, (2015)

\bibitem{Hosaka:2009}
K. Hosaka, S. A. Webster, A. Stannard, B. R. Walton, H. S. Margolis, and P. Gill, ``Frequency measurement of the ${^{2}S}_{1/2}\text{\ensuremath{-}}{^{2}F}_{7/2}$ electric octupole transition in a single ${^{171}\text{Y}\text{b}}^{+}$ ion,'' Phys. Rev. A {\bf 79}, 033403 (2009).

\bibitem{Huntemann:2012}
N. Huntemann, M. Okhapkin, B. Lipphardt, S. Weyers, Chr. Tamm, and E. Peik, ``High-Accuracy Optical Clock Based on the Octupole Transition in ${}^{171}$Yb$^{+}$,'' Phys. Rev. Lett. {\bf 108}, 090801 (2012).

\bibitem{Roberts:1999}
M. Roberts, P. Taylor, S. V. Gateva-Kostova, R. B. M. Clarke, W. R. C. Rowley, and P. Gill, ``Measurement of the $\LS{S}{1/2}$-$\LS{D}{5/2}$ clock transition in a single ${}^{171}\textrm{Yb}^+$ ion,'' Phys. Rev. A {\bf 60}, 2867 (1999).

\bibitem{Baldwin:2020}
C. H. Baldwin, B. J. Bjork, M. Foss-Feig, J. P. Gaebler, D. Hayes, M. G. Kokish, C. Langer, J. A. Sedlacek, D. Stack, and G. Vittorini, ``A high fidelity light-shift gate for clock-state qubits,'' Phys. Rev. A {\bf 103}, 012603 (2021).

\bibitem{Edmunds.2020b}
C. L. Edmunds, T. R. Tan, A. Milne, A. Singh, M. J. Biercuk, C. Hempel, ``Scalable hyperfine qubit state detection via electron shelving in the D$_{5/2}$ and F$_{7/2}$ manifolds in \Yb,'' submitted (2020).

\bibitem{Sugiyama:1999}
K. Sugiyama, ``Laser Cooling of Single ${}^{174}\textrm{Yb}^{+}$ Ions Stored in a {RF} Trap,'' Japanese Journal of Applied Physics {\bf 38}, 2141 (1999).

\bibitem{Jaua:2015}
Y.-Y. Jau, J. D. Hunker, and P. D. D. Schwindt, ``F-state quenching with $\textrm{CH}_4$ for buffer-gas cooled ${}^{171}\textrm{Yb}^+$ frequency standard,'' AIP Advances {\bf 5}, 117209 (2015).

\bibitem{Mulholland:2019}
S. Mulholland, H. A. Klein, G. P. Barwood, S. Donnellan, P. B. R. Nisbet-Jones, G. Huang, G. Walsh, P. E. G. Baird, and P. Gill, ``Compact laser system for a laser-cooled ytterbium ion microwave frequency standard,'' Rev. Sci. Ins. {\bf 90}, 033105 (2019).

\bibitem{Edmunds:2020}
C. L. Edmunds, C. Hempel, R. J. Harris, V. Frey, T. M. Stace, and M. J. Biercuk, ``Dynamically corrected gates suppressing spatiotemporal error correlations as measured by randomized benchmarking,'' Phys. Rev. Research {\bf 2}, 013156 (2020).

\bibitem{Ludlow:2015}
A. D. Ludlow, M. M. Boyd, J. Ye, E. Peik, and P. O. Schmidt, ``Optical atomic clocks,'' Rev. Mod. Phys. {\bf 87}, 637 (2015).

\bibitem{Meggers:1967}
W. F. Meggers, ``The Second Spectrum of Ytterbium (Yb II),'' Journal of research of the National Bureau of Standards. Section A, Physics and chemistry {\bf 71A}, 396 (1967).

\bibitem{MOGCatEye}
MOG Laboratories Model: CEL002

\bibitem{Thompson:2012}
D. J. Thompson and R. E. Scholten, ``Narrow linewidth tunable external cavity diode laser using wide bandwidth filter,'' Rev. Sci. Instrum. {\bf 83}, 023107 (2012).

\bibitem{Berends:1993}
R. W. Berends, E. H. Pinnington, B. Guo, and Q. Ji, ``Beam-laser lifetime measurements for four resonance levels of Yb {II},'' Journal of Physics B: Atomic, Molecular and Optical Physics {\bf 26}, L701 (1993).

\bibitem{Webster:1999}
S. A. Webster, P. E. Blythe, K. Hosaka, G. P. Barwood, P. Gill, ``Systematics shifts of the 467 nm electric octupole transition in 171 Yb$^+$,'' NPL Report CBTLM 31 (1999).

\bibitem{Taylor:1999}
P. Taylor, M. Roberts, G. M. Macfarlane, G. P. Barwood, W. R. C. Rowley, and P. Gill, ``Measurement of the infrared ${}^{2}{F}_{7/2}{\ensuremath{-}}^{2}{D}_{5/2}$ transition in a single ${}^{171}{\mathrm{Yb}}^{+}$ ion,'' Phys. Rev. A {\bf 60}, 2829 (1999).

\bibitem{Gill:1995}
P. Gill, H. A. Klein, A. P. Levick, M. Roberts, W. R. C. Rowley, and P. Taylor, ``Measurement of the $\LS{S}{1/2}$-$\LS{D}{5/2}$ 411-nm interval in laser-cooled trapped ${}^{171}\textrm{Yb}^{+}$ ions,'' Phys. Rev. A {\bf 52}, R909 (1995).

\bibitem{Fuerst:2020}
H. A. F{\"u}rst, C.-H. Yeh, D. Kalincev, A. P. Kulosa, L. S. Dreissen, R. Lange, E. Benkler, N. Huntemann, E. Peik, T. E. Mehlst{\"a}ubler, ``Coherent excitation of the highly forbidden electric octupole transition in ${}^{172}\textrm{Yb}^{+}$,'' Phys. Rev. Lett. {\bf 125}, 163001 (2020).

\bibitem{Martin:2002}
W. C. Martin and W. L. Wiese, ``Atomic, Molecular, and Optical Physics Handbook,'' National Institute of Standards and Technology. http://physics.nist.gov/Pubs/AtSpec/index.html.

\bibitem{Vanier:1989}
J. Vanier and C. Audoin, ``The Quantum Physics of Atomic Frequency Standards,'' Philadelphia : A. Hilger (1989).

\bibitem{Itano:2000}
Wayne M. Itano, ``External-Field Shifts of the ${}^{199}\textrm{Hg}^+$ Optical Frequency Standard,'' Journal of research of the National Institute of Standards and Technology {\bf 105}, 829 (2000).

\bibitem{Stone:2015}
N. J. Stone, ``Nuclear Magnetic Dipole and Electric Quadrupole Moments: Their Measurement and Tabulation as Accessible Data,'' J. Phys. Chem. Ref. Data {\bf 44}, 031215 (2015).

\bibitem{Gan:2018}
H. C. J. Gan, G. Maslennikov, K.-W. Tseng, T. R. Tan, R. Kaewuam, K. J. Arnold, D. Matsukevich, and M. D. Barrett, ``Oscillating-magnetic-field effects in high-precision metrology,'' Phys. Rev. A {\bf 98}, 032514 (2018).


\end{thebibliography}
\end{document}